\documentclass[12pt,aps,prd,floats,floatfix,showpacs,twoside,preprintnumbers,superscriptaddress,nofootinbib]{revtex4}
\usepackage[dvips]{graphicx,pstricks}
\usepackage{amsmath}
\usepackage{amssymb}
\usepackage{hyperref}
\usepackage{epsfig}
\usepackage{color}

\allowdisplaybreaks

\begin{document}
\date{\today}

\title{Nonlinear oscillations of compact stars in the vicinity of the maximum mass configuration}

\author{A.~Brillante}
\affiliation{Frankfurt Institute for Advanced Studies (FIAS), Ruth-Moufang-Stra{\ss}e 1, 60438 Frankfurt am Main, Germany}
\affiliation{Institut f\"{u}r Theoretische Physik, Johann Wolfgang Goethe-Universit\"{a}t, Max-von-Laue-Stra{\ss}e 1, 60438 Frankfurt am Main, Germany}

\author{I. N.~Mishustin}
\affiliation{Frankfurt Institute for Advanced Studies (FIAS), Ruth-Moufang-Stra{\ss}e 1, 60438 Frankfurt am Main, Germany}
\affiliation{Kurchatov Institute, Russian Research Center, Akademika Kurchatova Square, Moscow, 123182, Russia}

\begin{abstract}
We solve the dynamical GR equations for the spherically symmetric evolution of compact stars in the vicinity of the maximum mass, for which instability sets in according to linear perturbation theory. The calculations are done with the analytical Zeldovich-like EOS $P=a\left(\rho-\rho_0\right)$ and with the TM1 parametrisation of the RMF model. The initial configurations for the dynamical calculations are represented by spherical stars with equilibrium density profile, which are perturbed by either (i) an artificially added inward velocity field proportional to the radial coordinate, or (ii) a rarefaction corresponding to a static and expanded star. These configurations are evolved using a one-dimensional GR hydro code for ideal and barotropic fluids. Depending on the initial conditions we obtain either stable oscillations or the collapse to a black hole. The minimal amplitude of the perturbation, needed to trigger gravitational collapse is evaluated. The approximate independence of this energy on the type of perturbation is pointed out. At the threshold we find type I critical behaviour for all stellar models considered and discuss the dependence of the time scaling exponent on the baryon mass and EOS.
\end{abstract} 

\maketitle

\section{Introduction}
The research on stellar oscillations within the framework of GR goes back to the early works of Chandrasekhar and others from about half a century ago \cite{chandra1,chandra2,misner,harrison}. The most simple class of oscillations are \emph{radial} oscillations, which can be classified by a single index denoting the number of nodes within the star. Within the sequence of equilibrium configurations for a given equation of state (EOS), there is one for which the squared frequency of the fundamental mode crosses zero. It is usually identified as the critical star which corresponds to the configuration of maximum mass and maximum baryon number. In the mass-radius diagramm it separates the branch of \emph{perturbatively} unstable stars from the stable branch.
\\\\
\noindent The dynamics of gravitational collapse in GR was studied for the first time in \cite{shapiro-teukolsky}, see also references therein. The authors considered a polytropic equation of state $P=Kn^{\gamma}$ and a large number of different choices for $\gamma$, the total mass and the radius of the initial configurations, which were assumed to be at rest. The perturbations were imposed by streching the matter from its equilibrium radius to some larger radius and thereby obtaining an inner pressure depletion. Specifically, the authors asked the question, whether black holes could be formed out of stellar cores with a mass below the maximum mass for a given EOS. The conclusion was, that stars below the maximum mass are dynamically stable even for large perturbations, \emph{i.e.} they perform oscillations without evolving to a black hole.
\\\\
\noindent
Similar studies with a more realistic microphysical description of the EOS were performed in \cite{gourgoulhon,novak}. The initial configurations were neutron stars with an equilibrium density profile, to which an inward linear velocity field was added by hand. In contrast to \cite{shapiro-teukolsky} it was found that linearly stable neutron stars can collapse to a black hole, if the perturbation is large enough and the mass of the star exceeds a lower threshold depending on the EOS.
\\\\
\noindent Non-linear effects in the evolution of spherical stars have been studied in \cite{sperhake,sperhake1}. A detailed investigation of the nonlinear dynamics and its phenomenology was obtained in \cite{gabler}. It was found that mode pairs with an integer frequency ratio can efficiently interact and exchange energy. Furthermore, it was shown that linearly unstable stars can perform stable oscillations over a long period of time without collapsing to a black hole. The evolution of linearly super-critical stars in GR was also investigated in \cite{radice1}. They considered polytropes with a constant baryon mass and showed that the lifetime of the metastable solution along the evolutionary path exhibits the scaling relationship
\begin{equation}
\tau=-\sigma\;ln\;|p-p^{*}| + const
\label{typeI}
\end{equation}
, which defines the critical phenomenon of type I \cite{garcia}. Here, $\sigma$ denotes the \emph{time scaling exponent}, which controls the sharpness of the threshold behaviour near black hole formation. The \emph{metastable escape time} or \emph{lifetime of the intermediate state} $\tau$ measures how long the solution with parameter $p$ stays close to the critical solution with parameter $p^{*}$.
\\\\A thorough investigation of the critical collapse dynamics of low-mass neutron stars has been presented in \cite{noble1}, where the scaling behaviour of black hole formation was studied. This letter aims to extend the previous work in \cite{shapiro-teukolsky,gourgoulhon,novak,noble1} and to investigate radial oscillations of large amplitudes in the vicinity of the maximum-mass configuration, taking into account non-linear effects. Specifically, we want to address the question, how strong the perturbation should be, in order to traverse the barrier and cause a collapse to a black hole (see fig. 7). In contrast to previous work, we will identify configurations by their baryon number and therefore avoid the decomposition into background and perturbation, which is arbitrary to some extent.
\\\\
\noindent It is clear that the energy barrier, which separates stable neutron stars from black holes, must change smoothly in the vicinity of the maximal mass. In this letter we report on a systematic survey of this energy wall of massive compact stars and its dependence on EOS stiffness and baryon number, which is lacking in previously published work. The envisioned collapse may be triggered by mass accretion in binary systems or in a supernova explosion, where the expelled layers fail to reach the escape velocity and fall back onto the proto-neutron star.
\\\\
\section{Dynamical equations}

In this section we consider non-linear radial oscillations of compact stars. We keep the assumption of spherical symmetry, but allow for perturbations of arbitrary amplitude. We assume that there are no gradients in the specific entropy (homentropic fluid). In this case the equation for baryon number conservation is not linearly independent from the equation for energy-momentum conservation. The equations governing the dynamics are Einstein's equation and the energy-momentum conservation equation.\footnote{Here and below we use geometrical units, $G=c=km=1$.}
\begin{align}
&G^{\mu\nu}=8\pi T^{\mu\nu}\label{einstein}\\
&T^{\mu\nu}_{\;\;\;\;;\nu}=0\label{emc}
\end{align}
We adopt the Schwarzschild-like parametrisation of the metric 
\begin{align}
ds^2=-e^{2\Phi(t,r)}dt^2+e^{2\Lambda(t,r)}dr^2+r^2d\theta^2+r^2sin^2\theta d\phi^2
\end{align}
with time-dependent metric functions $\Phi$ and $\Lambda$. The normalisation of the 4-velocity yields
\begin{align}
u^0=\left[e^{2\Lambda-2\Phi}{u^1}^2+e^{-2\Phi}\right]^\frac{1}{2}\label{4-velocity}
\end{align}
a relationship between $u^0$ and $u^1$ (henceforth denoted by $u$). It should be stressed that our choice of the metric involves a coordinate singularity at the Schwarzschild radius and thus we cannot follow the motion of matter trough the event horizon. We introduce the rescaled radial coordinate $\eta=r/R(t)$, where $R(t)$ is the time-dependent stellar radius. Then eq. \eqref{einstein} leads to the two constraint equations for the metric functions
\begin{align}
\Phi,_{\eta}&=4\pi R^2\eta\left(\rho+P\right)e^{4\Lambda}u^2+4\pi R^2\eta P e^{2\Lambda}+\frac{e^{2\Lambda}-1}{2\eta}\label{constrain1}\\
\Lambda,_{\eta}&=4\pi R^2\eta\left(\rho+P\right)e^{4\Lambda}u^2+4\pi R^2\eta \rho e^{2\Lambda}+\frac{1-e^{2\Lambda}}{2\eta}\label{constrain2}
\end{align}
where "$,$" denotes a partial derivative, $\rho$ the total energy density (including rest mass) and $P$ the pressure. Adopting the energy-momentum tensor for an ideal fluid, eq. \eqref{emc} provides a hyperbolic set of two evolution equations for the pressure and fluid 4-velocity
\begin{align}
P,_{t}&+c_{11}P,_{\eta}+c_{12}u,_{\eta}+c_{13}=0\label{evolve1}\\
u,_{t}&+c_{21}P,_{\eta}+c_{11}u,_{\eta}+c_{23}=0\label{evolve2}
\end{align}
Here we have introduced the notations:
\begin{align}
c_{11}&=\frac{\left(\frac{\partial \rho}{\partial P}-1\right)e^{2\Phi}u^0u}{R\xi}-\frac{\dot{R}\eta}{R}\\
c_{12}&=\frac{\rho+P}{Ru^0\xi}\\
c_{13}&=\frac{e^{\Phi}u^0u}{2r\left(1+e^{2\Lambda}{u}^2\right)\xi}\bigg[5e^{\Phi}\left(\rho+P\right)+e^{2\Lambda+\Phi}\left(\rho+P\right)\big(-1-8\pi Pr^2+4{u}^2\big)\bigg]\\
c_{21}&=\frac{\frac{\partial\rho}{\partial P}e^{2\Phi-2\Lambda}u^0}{R\left(\rho+P\right)\xi}\\
c_{23}&=\frac{-e^{\Phi-2\Lambda}u^0}{2r\xi}\bigg[\frac{\partial\rho}{\partial P}\Big(e^{\Phi}+e^{2\Lambda+\Phi}\left(-1-8\pi Pr^2\right)\Big)+4e^{2\Lambda+\Phi}{u}^2\bigg]
\end{align}
and 
\begin{align}
\xi&=\frac{\partial\rho}{\partial P}-e^{2\Lambda}{u}^2+\frac{\partial\rho}{\partial P}e^{2\Lambda}{u}^2\text{.}
\end{align}
The set of eqs. for the five unknown variables $P$, $\rho$, $u^{1}$, $\Phi$, $\Lambda$ is given by eqs. \eqref{constrain1}\eqref{constrain2}\eqref{evolve1}\eqref{evolve2} together with the equation of state. The enclosed baryon number $N_B$ obeys the differential equation
\begin{equation}
N_{B\;,\eta}=4\pi R^3\eta n e^{\Phi+\Lambda} u^0\label{constrain_baryon_number}\text{,}
\end{equation}
where $n$ denotes the baryon number density. The evolution of the star radius is obtained from the equation
\begin{align}
\dot{R}=\frac{u(R)}{u^0(R)}\label{evolution-radius}\text{.}
\end{align}
\\
Equations \eqref{evolve1} and \eqref{evolve2} are equivalent to eqs. (5.85) and (5.86) in \cite{sperhake1}, as can be demonstrated by transforming the radial coordinate from $\eta$ to r. It is easy to show that for the stationary limit eq. \eqref{evolve1} is trivially fullfilled, whereas eq. \eqref{evolve2} leads to the well known TOV equation of relativistic stellar structure. The eigenmode equation for radial modes can be obtained by perturbing eq. \eqref{evolve2} around equilibrium \cite{chandra1,chandra2,brillante} and imposing harmonic time-dependence.
\\\\

\section{Equation of state}
\subsection{Zeldovich EOS}
In our calculations we have used two equations of state. The first one is the Generalised Zeldovich (GZ) EOS $P=a\left(\rho-\rho_0\right)$, which for $a=1$ yields the {\it causal-limit} EOS proposed by Ya. Zeldovich \cite{zeldovich} and which for $a=1/3$ coincides with the MIT bag model\cite{chodos}. This simple parametrisation is able to qualitatively reproduce the bulk properties of neutron stars. While $a$ controls the stiffness and the sound velocity, the parameter $\rho_0$ introduces a dimensional scale in the EOS. Different choices for $a$ and $\rho_0$ allow us to consider families with different maximum mass or radius. Interestingly, the compactness ($=\,2\,M/R$) of the maximum mass or maximum radius configuration does not depend on $\rho_0$. In fig. 1 we plot the mass of the maximum-mass configuration for the parameter space given by $a$ and $\rho_0$. The high mass stars in the bottom-right corner correspond to low central densities, while the low mass stars in the top-left corner correspond to large central densities. In fig. 2 we show the radii of the same stellar models as in fig. 1. The baryon number density $n$ is related to the energy density $\rho$ by $n/n_0=\left[a\left(\rho+P\right)\right]^{1/\left(1+a\right)}$. Here $n_0$ is a free scaling variable, which needs to be found from a specific matter model. We note that the baryon number ratio of two stellar models does not depend on the choice of $n_0$. This equation solves the well known identity for the baryon chemical potential $\mu_B$ of cold neutron star matter \cite{haenselbook}.
\begin{figure}[!t]
\begin{center}
\includegraphics[width=0.8\linewidth]{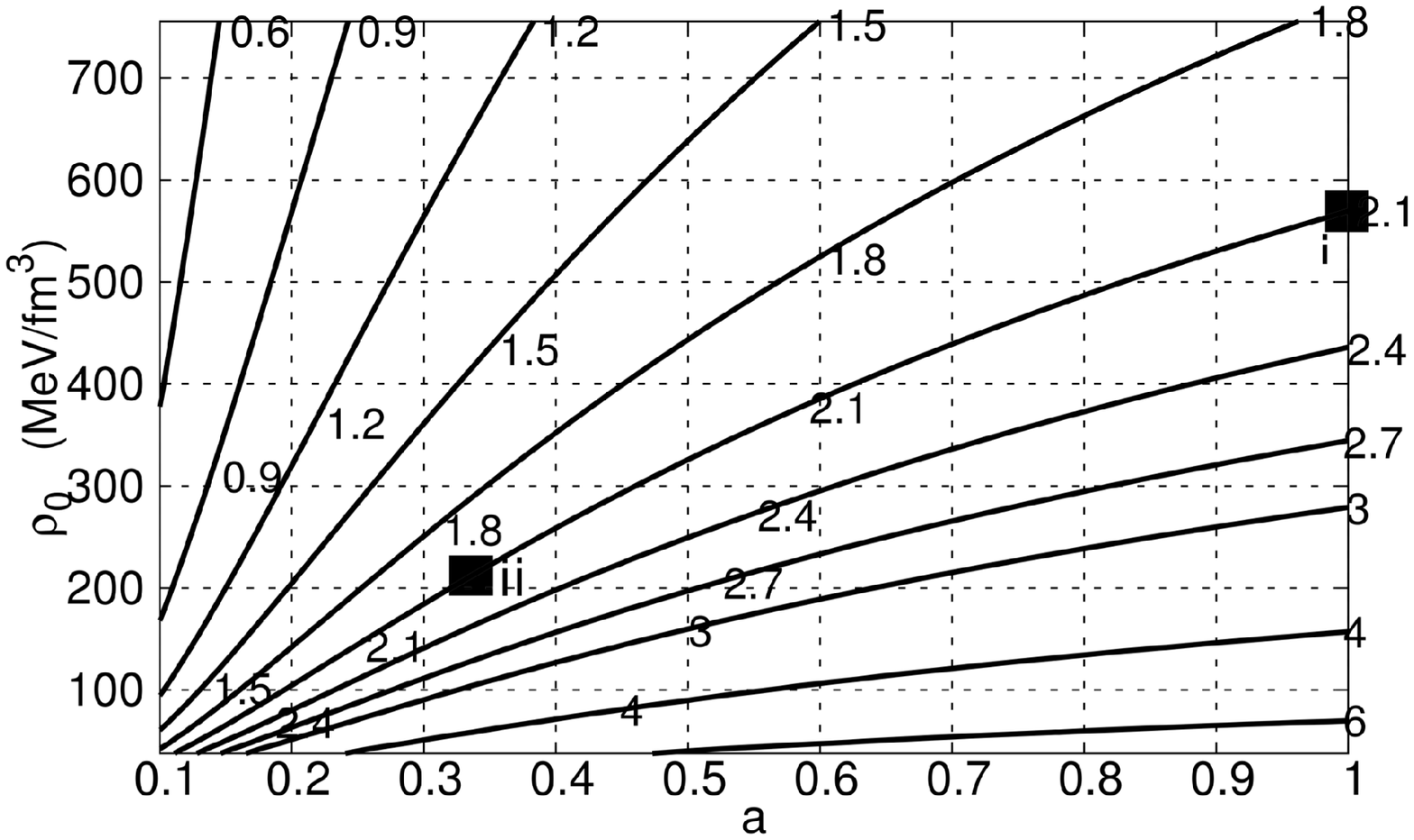}
\end{center}
\label{fig:p11}
\caption{Maximum mass $M$ of compact stars with the generalised Zeldovich EOS $P=a\left(\rho-\rho_0\right)$ as function of $a$ and $\rho_0$. The lines indicate configurations of constant $M/M_{Sun}$. The choices of $a$, $\rho_0$ for the simulations are indicated by solid squares.}
\begin{center}
\includegraphics[width=0.8\linewidth]{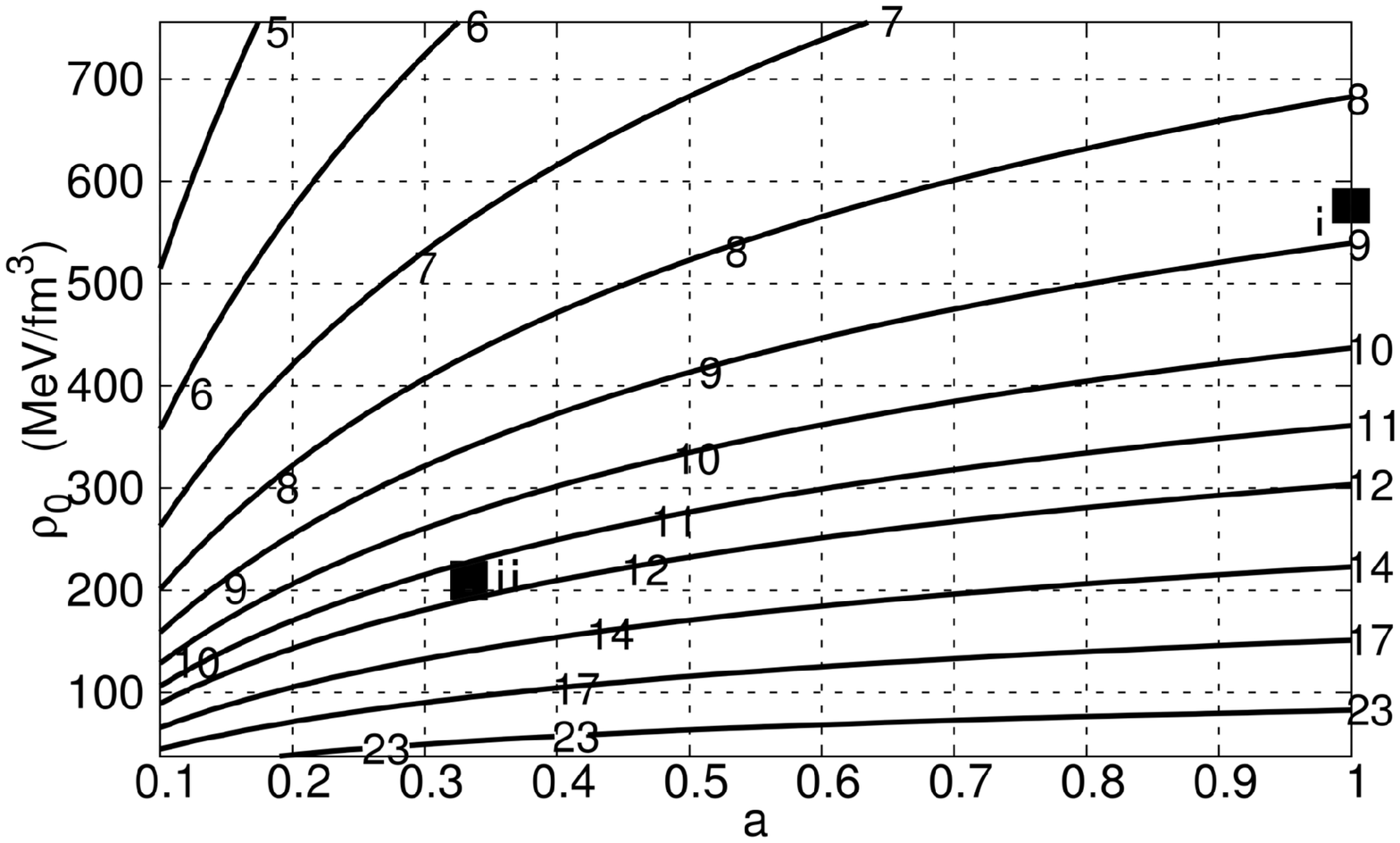}
\end{center}
\label{fig:p2}
\caption{Radius $R$ of the maximum mass configuration of compact stars with the generalised Zeldovich EOS $P=a\left(\rho-\rho_0\right)$ as function of $a$ and $\rho_0$. The lines indicate configurations of constant $R/km$.}
\end{figure}

\begin{equation}
\frac{dn}{d\rho}={\mu_B}^{-1}=\frac{n}{\rho+P}
\end{equation}
\subsection{RMF EOS}
For a realistic description of neutron stars, the relativistic mean field models are often employed \cite{reinhard}. They provide an effective and relativistically covariant description of dense baryonic systems. The parameters are fitted to the known properties of nuclear matter at saturation. As the second EOS we use the TM1 parameter set \cite{sugahara}, which leads to a maximum mass of about $2.2 M_{\odot}$.

\section{Results}
We consider 3 different classes of star configurations based on the above-mentioned EOS, namely\\
(i) the GZ model with $a=1$, $\rho_0=7.5367\;10^{-4} km^{-2}$\\
(ii) the GZ model with $a=1/3$, $\rho_0=2.7782\;10^{-4} km^{-2}$\\
(iii) the TM1 EOS based on the relativistic mean field model.\\
The parameters in the first two cases are chosen to yield a maximum mass of precisely $2.1 M_{\odot}$ (see fig. 1), whereas the TM1 EOS yields a maximum mass of approximately $2.2 M_{\odot}$. For each EOS we select 10 models based on the baryon number of the maximum mass configuration $N_{B,max}$, namely configurations with a baryon number $N_B=\left(1-k\,10^{-2}\right)N_{B,max}$ with $k=1,2..10$.
\subsection{Initialisation}
\begin{figure}[!t]
    \begin{center}
\vspace*{-0.9cm}
        \includegraphics[width=0.83\linewidth]{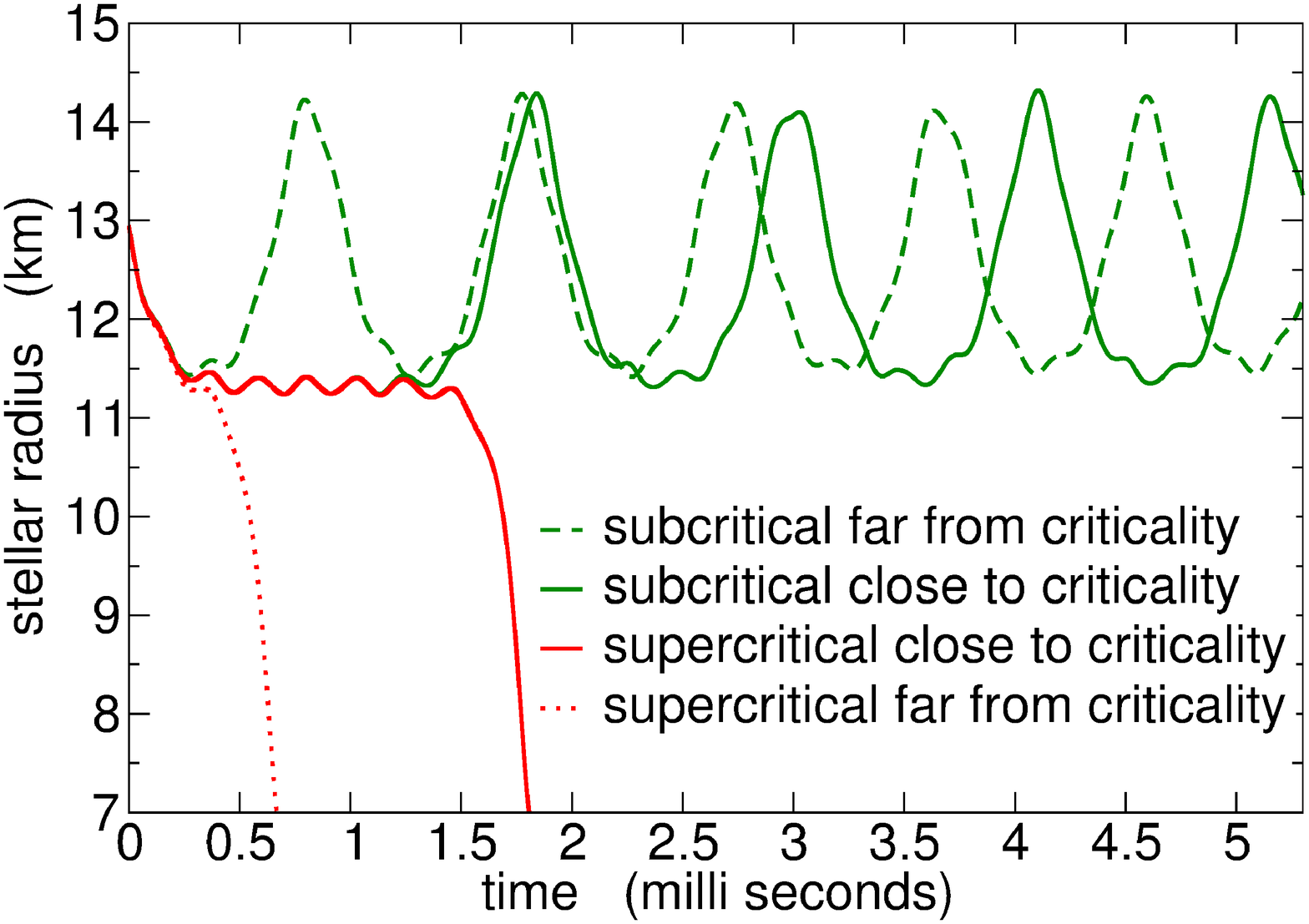}
        \setcounter{figure}{2}
        \label{fig:p11}
\vspace*{-0.5cm}
        \caption{The time evolution of the stellar radius for the TM1 EOS with a baryon number $4\,\%$ below the maximum value. 4 trajectories corresponding to a different strength of the velocity field induced perturbation are shown. The velocity profile belonging to the red solid line is shown in fig. 5.}
        \includegraphics[width=0.98\linewidth]{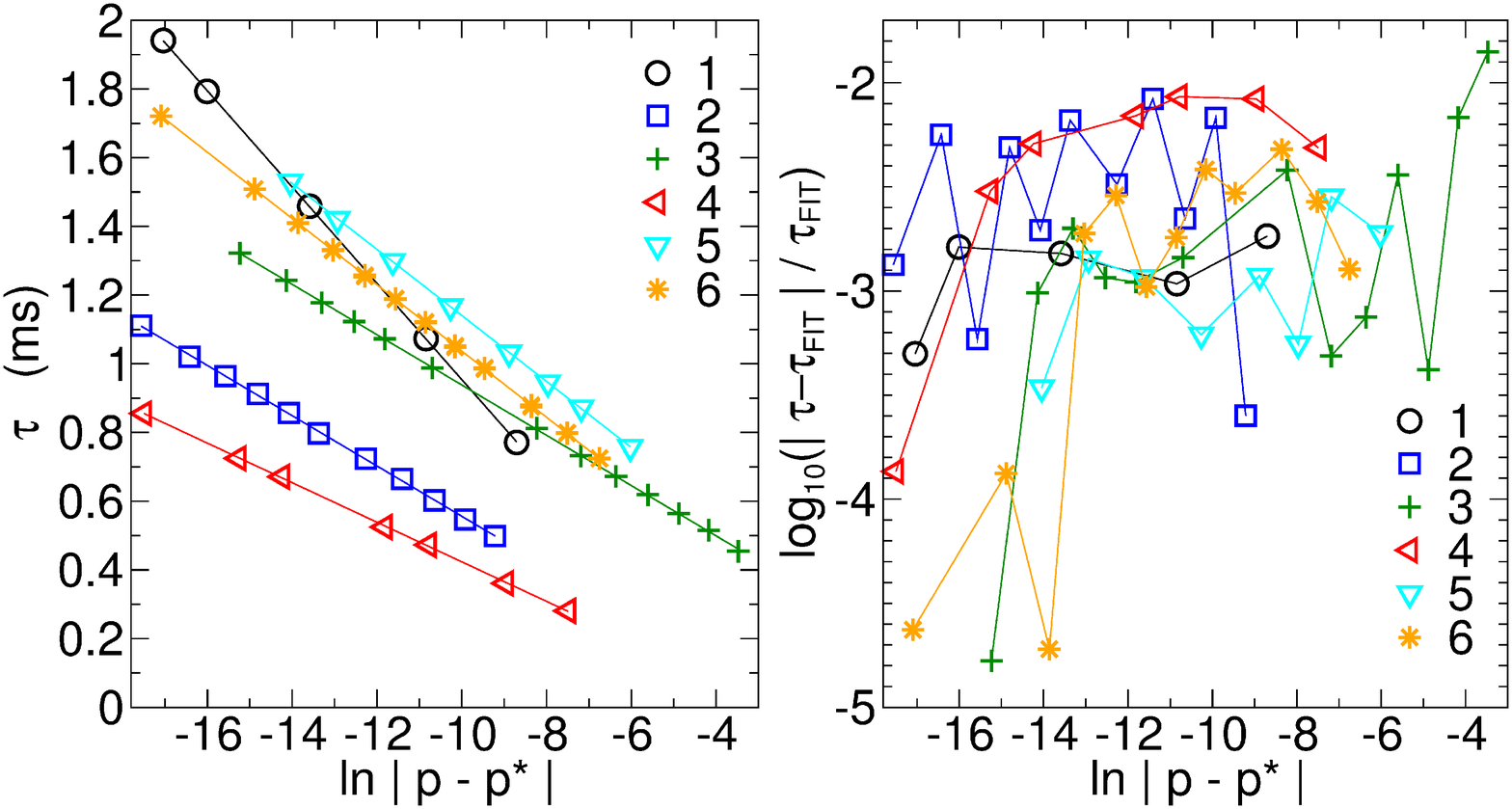}
\vspace*{-0.6cm}
        \setcounter{figure}{3}
        \label{fig:p2}
        \caption{The metastable escape time (left) and its relative error with respect to the linear fit (right) as function of initial data. The legend indicates the specific stellar models, namely 1: (iii,1,v), 2: (iii,10,v), 3: (iii,10,r), 4: (i,10,v), 5: (iii,4,r), 6: (ii,4,r).}
    \end{center}
\end{figure}

\begin{figure}
    \begin{center}
\vspace*{-1.3cm}
        \includegraphics[width=0.9\linewidth]{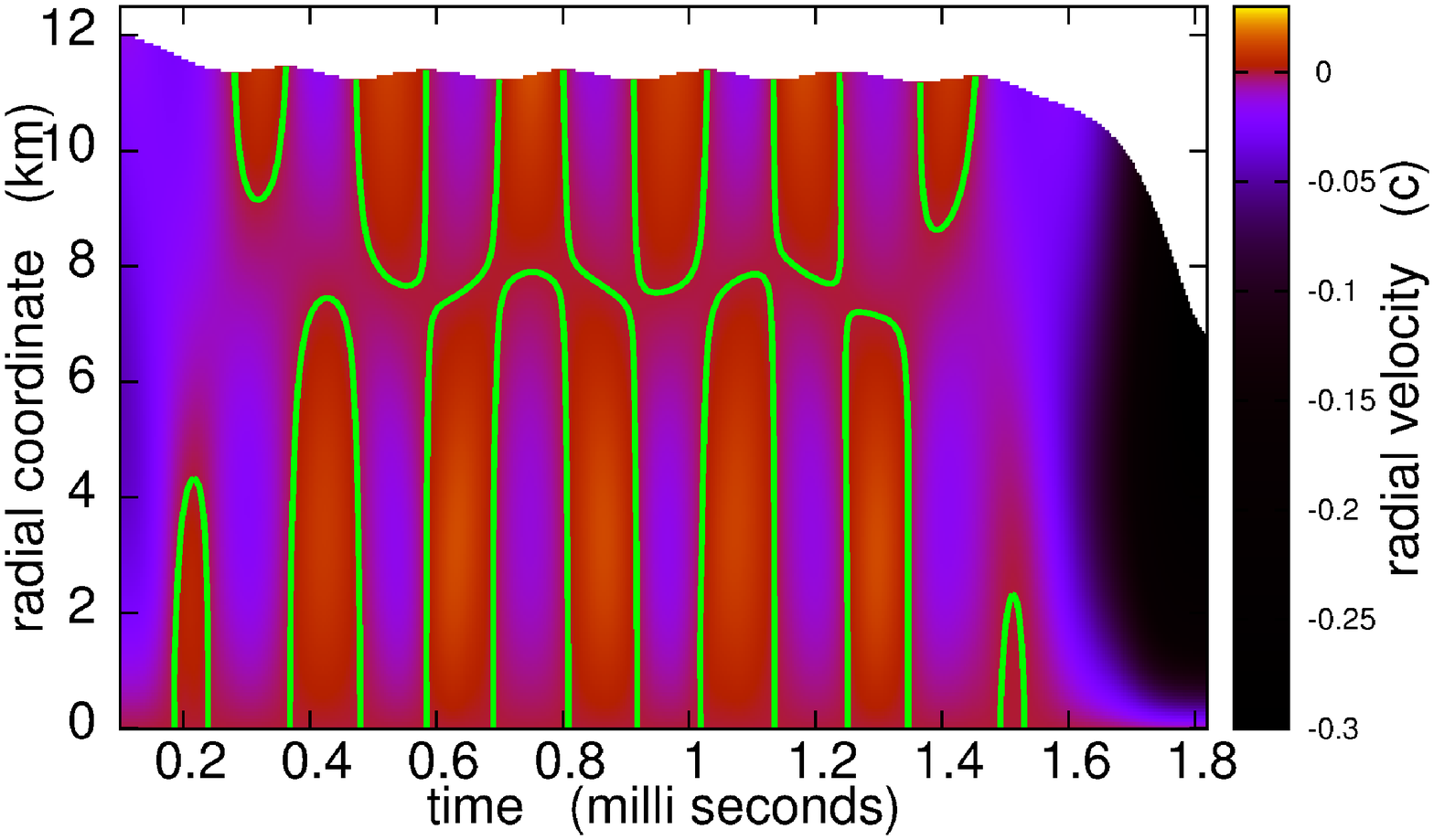}
        \setcounter{figure}{4}
        \label{fig:p11}
\vspace*{-0.6cm}
        \caption{The radial velocity $u$ as a function of time and radius for the slightly supercritical solution shown in fig. 3. The green contours denote zero velocity points and separate infalling matter from expanding matter.}
        \includegraphics[width=0.9\linewidth]{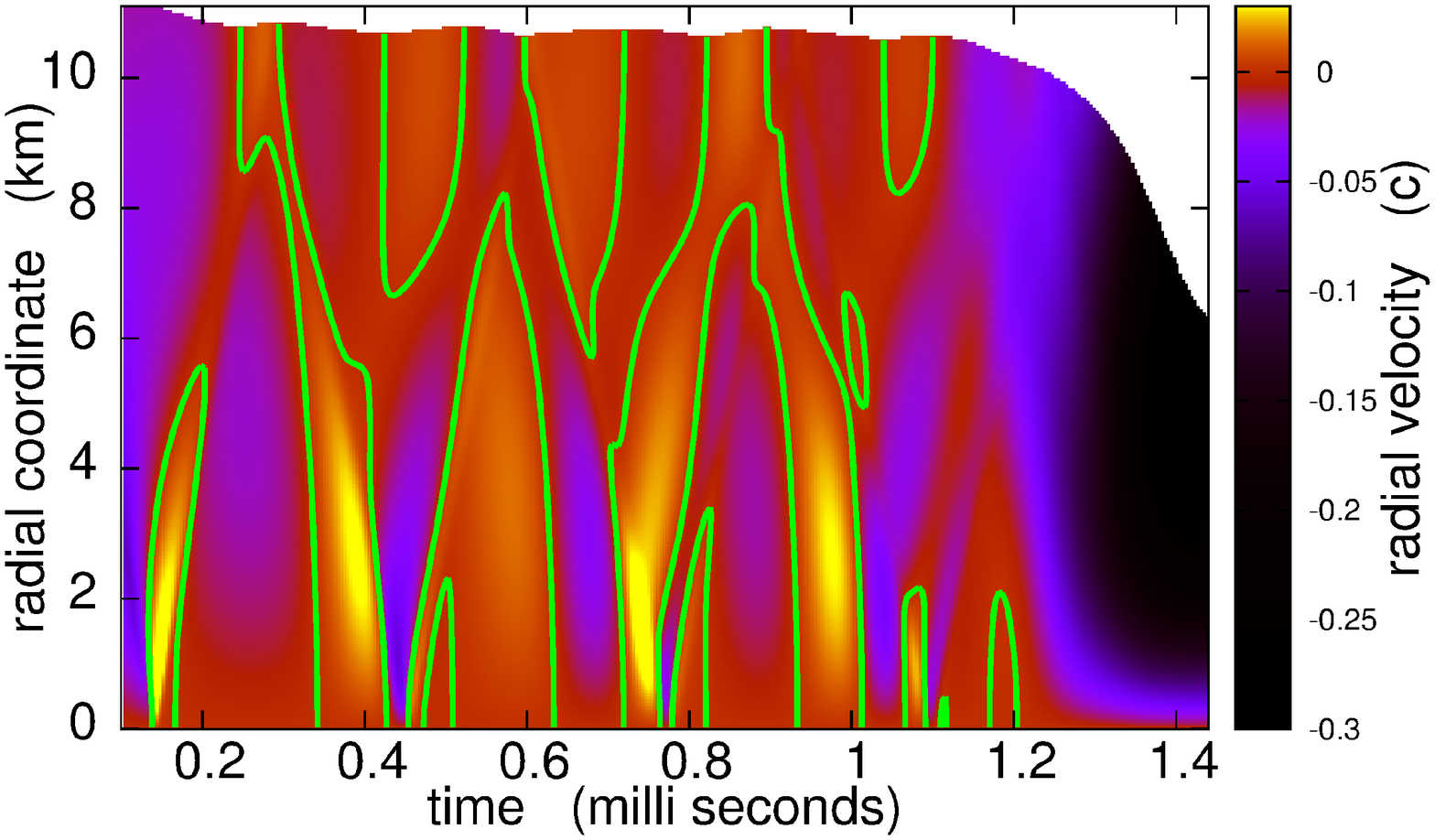}
        \setcounter{figure}{5}
        \label{fig:p2}
\vspace*{-0.6cm}
        \caption{The radial velocity $u$ as a function of time and radius for a slightly supercritical solution belonging to the case (ii,4,v). The green contours denote static fluid elements.}
    \end{center}
\end{figure}

We consider two different types of initial perturbations preserving the baryon number, namely the velocity field induced perturbation and the rarefaction induced perturbation.\\
\emph{Velocity field induced perturbation}: Once $N_B$ is specified, we choose the central energy density, which leads to a stable equilibrium solution with baryon number $N_B$. We then impose an inward velocity field $u(\eta)=-\eta\;u_{max}$, where $u_{max}$ is the surface velocity. To preserve $N_B$ according to eq. \eqref{constrain_baryon_number}, the density and pressure profiles of the TOV solutions have to be adapted. We therefore select a radius independent $x$ and redefine the pressure according to $P_{new}(\eta)=P_{old}(\eta)\,\left(1+x\right)$. Since the EOS is barotropic, this procedure also determines the energy density. This rescaling allows to preserve $N_B$, while considering perturbations of different strength. Note that this procedure does not affect the stellar radius.\\
\emph{Rarefaction induced perturbation}: Once $N_B$ is selected, we again choose the central energy density, which leads to a stable equilibrium solution with baryon number $N_B$. The stellar radius is then artificially increased, while $\rho\left(\eta\right)$ and $P\left(\eta\right)$ are held constant. In order to preserve $N_B$ according to eq. \eqref{constrain_baryon_number}, we apply the same rescaling of density and pressure as described above. In contrast to above, here the initial stellar radius is not preserved and the perturbed star is initially at rest. Despite the names we attach to the two distinct kinds of initial perturbations, both involve a rarefaction, which in the first case however is considerably weaker. Henceforth we use tripels (EOS type \{i,ii,iii\}, baryon deficit \{1,2..10\}, perturbation type \{v,r\}) to label each of the 60 cases considered in the present study.\\\\
Figure 3 shows the evolution of the stellar radius for different magnitudes of the velocity induced perturbations. If the excitation energy is subcritical, the star performs oscillations of the fundamental mode, together with overtones. As the energy is shifted closer to criticality, the frequency of the fundamental oscillation decreases. We also notice that the shape of the oscillations deviates more and more from the harmonic curve. The closer the star is to the critical regime, the longer are delays at the compression phase. In other words, the closeness to criticality manifests itself in a long \emph{frustrated phase}, which is characterised by very small amplitude oscillations and can extend over several milli seconds. If the excitation energy is supercritical, the star evolves to a black hole.\\\\
For each case the bisection method is used to tune the strength of the initial perturbation towards criticality at the threshold between secular oscillations and gravitational collapse. Figure 4 shows the escape time $\tau$ and its deviation from a linear fit as function of the closeness of initial data to criticality. The parameter $p$ has been identified with $u_{max}$ for velocity induced perturbations and with the initial excess in radius $\Delta R$ otherwise. Here, $\tau$ has been identified as the length of the time interval during which the stellar radius deviates by less than 2\% from the solution obtained at the last bisection step.\footnote{Only subcritical evolutions have been used for the scaling relationship. The inclusion of supercritical ones does not change the picture.} For any value of $p^*$ a unique time scaling relation ensues. To determine the critical $p^*$ in the sense of eq. \eqref{typeI} we minimise the error of the linear fit of the ensuing relation with respect to a free $p^*$.\footnote{In practice this yields a value very close to the last bisection step.} The energy associated with this perturbation is denoted as \emph{critical excitation energy} or short \emph{critical energy}. It is the energy difference between the critical solution and the stable equilibrium solution of the same baryon number and can be calculated from eq. \eqref{constrain2}. \\\\The result of our analysis is shown in fig. 7, where we plot the critical excitation energy as function of the baryon number in units of the maximum baryon number of the selected EOS. The critical excitation energy decreases with increasing baryon number and the curves converge as the distance to the maximum mass configuration diminishes. Furthermore, regardless of the baryon number considered, the soft GZ case (case ii) exhibits a smaller critical energy than the TM1 case (case iii), which itself lies below the stiff GZ case (case i). The dependence on the type of perturbation is remarkably weak. For the case of the velocity induced perturbation we find an invariance of the critical energy with respect to the sign of the velocity, i.e. the result is exactly the same for initially ingoing and outgoing flow.
\begin{figure}[!t]
    \begin{center}
\vspace*{-0.9cm}
        \includegraphics[width=0.83\linewidth]{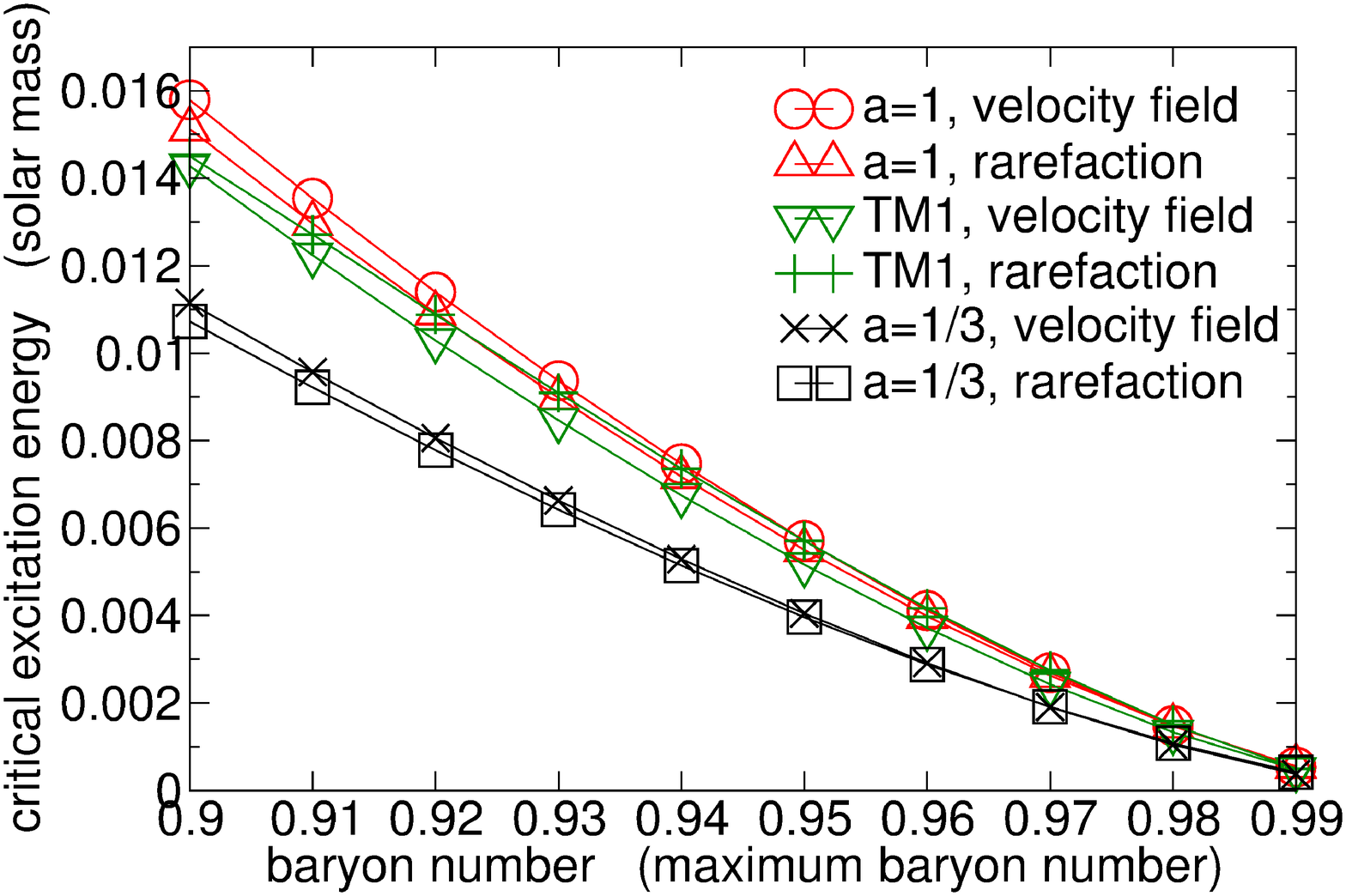}
        \setcounter{figure}{6}
        \label{fig:p2}
\vspace*{-0.5cm}
        \caption{The critical excitation energy as function of the baryon number. The unit on the x-axis is the maximum baryon number of the considered EOS, which is arbitrary for case (i) (see free scaling variable above), $3.4773\;10^{57}$ for case (ii) and $3.0704\;10^{57}$ for case (iii).}
        \includegraphics[width=0.83\linewidth]{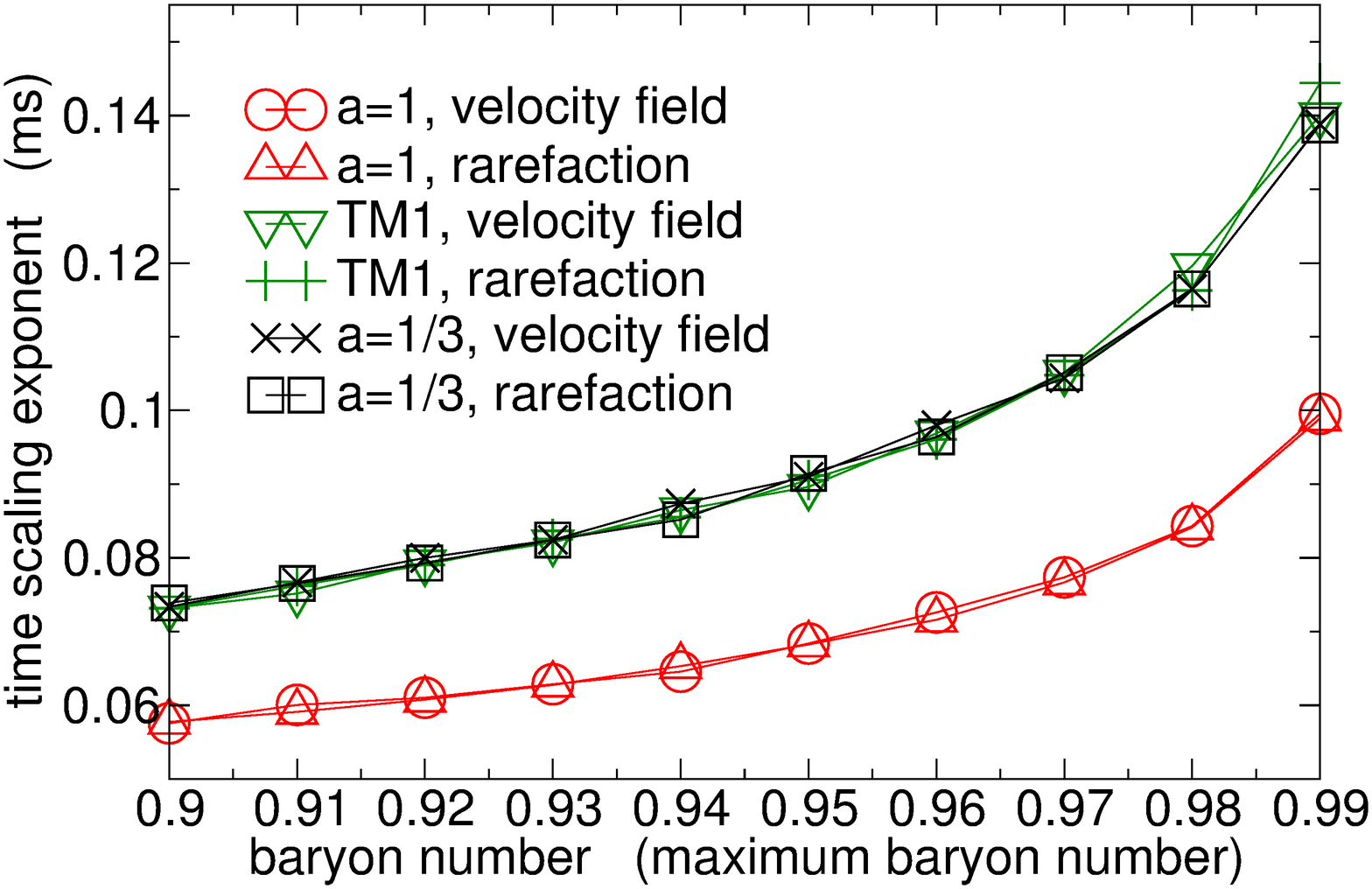}
        \setcounter{figure}{7}
        \label{fig:p2}
\vspace*{-0.5cm}
        \caption{The time scaling exponent (see eq. \eqref{typeI}) as function of the baryon number for the several considered EOS and perturbation types.}
    \end{center}
\end{figure}
\\\\
Next, we study the behaviour of the time scaling exponent $\sigma$ as function of the baryon number. We recall from eq. \eqref{typeI} that $\sigma$ is defined as the negative slope of the curve, which relates the lifetime of the intermediate state with $ln\;|p-p^*|$ for any one-dimensional parametetrisation p of initial data. Within the parameter space of initial data, $\sigma$ controls the width of the neighbourhood around criticality, for which threshold behaviour is significant. In practice, we have obtained $\sigma$ as the negative slope of the linear fit for the subset of subcritical evolutions.\\\\
The result of our analysis is shown in fig.8. The scaling exponent steadily increases with baryon number and gains about 70\% within the band of rest masses considered. There is no noticeable dependence on the perturbation type. Remarkably, we find identical behaviour for the EOS cases (ii) and (iii), i.e. strange stars with the simple MIT bag model and hadronic stars with the TM1 EOS exhibit exactly the same time scaling exponent. In contrast, for case (i) we find a reduction of $\sigma$ by about 30\%. Our reported values consistently exceed those reported in \cite{wan} for polytropic "Gaussian packet systems" by a factor of 1.2 to 2.5. We find the same dependence on baryon number, even though our result seems to support a convex dependence and we cannot confirm their flattening at high baryon masses.

\section{Conclusion}
In this work we have studied the dynamical evolution of nonlinear perturbations on spherical configurations with a baryon mass above 90\% of the maximal value of the respective EOS. Numerical simulations in full GR were performed with no other restrictions, than spherical symmetry and ideal hydrodynamics. The calculations were done for 3 different EOS and 10 different values of the stellar rest mass. A bisection strategy was invoked to determine the minimal energy for each EOS and baryon mass, which leads to black hole formation. This energy turns out to be more than an order of magnitude below the mass difference to the maximum mass configuration of the considered EOS. The critical energy increases with decreasing baryon number, i.e. the stars on the stable branch of the mass-radius diagram are \emph{stable to a different extent}. Furthermore its dependence on the type of perturbation is weak, but we have found a noticeable dependence on the EOS. Compact stars with stiff equations of state tend to be more resistent to collapse than their soft counterparts. In particular, strange quark stars should collapse easier to black holes than their hadronic counterparts.
\\\\
Furthermore, we have analysed the threshold behaviour close to black hole formation and found excellent agreement with the scaling relationship \eqref{typeI} for all the models considered. We therefore conjecture, that type I critical behaviour is a generic feature of high-mass compact stars. Remarkably, we have found universal scaling behaviour for hadronic and strange quark stars, independent of the perturbation type.  On the other hand, the weak dependence of the excitation energy on the specific type of initial perturbation suggests that the critical solution is \emph{not} universal with respect to different families of initial data and hence the initialisation of the perturbations in GR-critical collapse scenarios deserves particular attention. Whether this is due to an inconsistent treatment of shocks and thermodynamics in our simulations of the lower mass stars or a non-universality of the critical solution, we wish to explore in future work.

\acknowledgments
We thank J. Schaffner-Bielich, S. Schramm and R. P. Negreiros for fruitful discussions. This work was partially supported by the Hessian LOEWE initiative through the Helmholtz International Center for FAIR and the Helmholtz Graduate School for Hadron and Ion Research. I. M. acknoledges partial support from the grant NSH-215.2012.2 (Russia).

\end{document}